\documentclass[9pt,twocolumn,twoside]{pnas-new}

\setboolean{displaywatermark}{false}
\setboolean{displaycopyright}{false} 
\templatetype{pnasresearcharticle} 

\title{
 Causality inference in stochastic systems from neurons to currencies: Profiting from small sample size}

\author[a,b]{Danh-Tai Hoang}
\author[c,d,e]{Juyong Song} 
\author[a,1]{Vipul Periwal}
\author[c,d,f,1]{Junghyo Jo}

\affil[a]{Laboratory of Biological Modeling, National Institute of Diabetes and Digestive and Kidney Diseases, National Institutes of Health, Bethesda, Maryland 20892, USA}
\affil[b]{Department of Natural Sciences, Quang Binh University, Dong Hoi, Quang Binh 510000, Vietnam}
\affil[c]{Asia Pacific Center for Theoretical Physics, Pohang, Gyeongbuk 37673, Korea}
\affil[d]{Department of Physics, Pohang University of Science and Technology, Pohang, Gyeongbuk 37673, Korea}
\affil[e]{The Abdus Salam International Centre for Theoretical Physics, Strada Costiera 11, 34014 Trieste, Italy}
\affil[f]{School of Computational Sciences, Korea Institute for Advanced Study, Seoul 02455, Korea}

\leadauthor{Hoang} 

\significancestatement{Stochasticity is a dominant feature of natural dynamical phenomena. Maximum likelihood estimation (MLE) is standard for stochastic model inference, but MLE converges to the true parameter values only in the large sample limit. When the data is insufficient, as in neuronal dynamics, or the observed stochastic dynamics are modulated by time-varying deterministic trends, as in financial markets, alternatives to MLE are required. We use the mathematical formalism of statistical physics to define a free energy of data that measures the plausibility of observed configurations. Minimizing this free energy without resorting to gradient descent provides a computationally effective way to infer system couplings from small sets of observations, even including higher-order interactions. We demonstrate applications ranging from biological to financial networks.}

\authorcontributions{Please provide details of author contributions here.}
\authordeclaration{Please declare any conflict of interest here.}
\correspondingauthor{\textsuperscript{1}To whom correspondence should be addressed. E-mail: vipulp@mail.nih.gov or jojunghyo@kias.re.kr}

\keywords{Network reconstruction $|$ Inverse problem $|$ Time series $|$ Kinetic Ising model $|$ neuronal network $|$ Currency exchange} 

\begin{abstract} 
Success in modeling complex phenomena such as human perception hinges critically on the availability of data and computational power.  Significant progress has been made in modeling  such phenomena using probabilistic methods, particularly in image analysis and speech recognition.
Maximum Likelihood Estimation (MLE) combined with Bayesian model selection is the basis of much of this progress, as MLE converges to the true model with copious data. In the sciences, large enough datasets are rarae aves, so alternatives to MLE must be developed for small sample size. We introduce a data-driven statistical physics approach to model inference based on minimizing a free energy of data and show superior model recovery for small sample sizes. We demonstrate coupling strength inference in non-equilibrium kinetic Ising models, including in the difficult large coupling variability regime, and show scaling to systems of arbitrary size.
As applications, we infer a functional connectivity network in the salamander retina and a currency exchange rate network from time-series data of neuronal spiking and currency exchange rates, respectively. Accurate small sample size inference is critical for devising a profitable currency hedging strategy. 
\end{abstract}

\dates{This manuscript was compiled on \today}

\begin{document}

\verticaladjustment{-2pt}

\maketitle
\ifthenelse{\boolean{shortarticle}}{\ifthenelse{\boolean{singlecolumn}}{\abscontentformatted}{\abscontent}}{}

An explosion in data availability in recent years has ushered in a new era of data-driven research for natural and social sciences.
Identifying systems dynamics from observed data, e.g. biochemical reactions~\cite{Klimovskaia2016}, gene expression measurements~\cite{Bar2012}, neuronal or brain region activities~\cite{dombeck2007imaging,Schneidman2006,Nguyen2016, Bernal-Casas2017}, and population dynamics~\cite{Sugihara2012}, is of fundamental interest in science~\cite{Schmidt2009, Brunton2016, Yair2017,Nguyen2017, Natale2017}.  For complex phenomena, such as human perception, modeling system dynamics in a probabilisitic framework became possible with the advent of inexpensive computational resources, and has led to great progress in the last 25 years. 
Regardless of whether stochasticity is inherent in the system, or only apparent due to partial observability~\cite{Raj2008}, many stochastic processes have been analyzed by autoregressive-moving-average models~\cite{Hamilton1994} or probabilistic directed acyclic graphical models, often termed Bayesian networks~\cite{Friedman2004}.

The structure of such dynamic processes is often unknown and, in the social sciences in particular, there may be no underlying fundamental theory to delineate possible models.
Thus, a universal model-free data-driven approach has merit for the inference of  models from time-series data~\cite{Janes2006}.
Machine learning using recurrent neuronal networks is such an approach~\cite{Connor1994}, but it usually requires a large amount of training data and is computationally intensive.
Given time series of $N$ variables, network inference rapidly becomes too complex with increasing $N.$
Even considering only pair-wise interactions requires determining $N^2$ parameters and demands $L \ge N^2$  samples.
Including higher-order interactions leads to an exponential increase in the number of model parameters, and a concomitant increase in sample size.
In scientific contexts, however, we often encounter the case that data generated from experiments are not big enough to reconstruct the interaction network for a given system.
Theorists contend with the computational difficulties of inferring large systems by positing properties such as sparsity of interactions or specifying distributions of couplings, usually with scant experimental support. 

Maximum Likelihood  Estimation (MLE) is the gold standard for stochastic model parameter inference, as it converges to the true model parameters in the limit of large sample size. On the other hand, MLE is limited by the fact that the likelihood equations are specific to a given estimation problem, that the numerical estimation is usually non-trivial, and most importantly, MLE can be heavily biased for small samples where the optimality properties of MLE may not apply. MLE can also be sensitive to the choice of starting values~\cite{nistwebsite}. 

According to the Rao-Blackwell theorem~\cite{Rao1945, Blackwell1947}, 
the conditional expected value of an estimator given a sufficient statistic is another estimator that is at least as good, and this result applies to MLE estimators as well. The  Rao-Blackwell result usually applies for sufficient and complete statistics, and leads to an idempotent improvement, in other words, the improvement requires no iteration. However, for our small sample size purposes, more apropos is the recent result of Galili and Meilijson~\cite{Galili2016}, which suggests that a Rao-Blackwell--type iterative improvement of a parameter estimator is  worth investigating. 

Statistical physics is often used for model inference~\cite{Sohl-Dickstein2011, Decelle2014}, 
but, in fact, for small sample sizes, the observed configurations of the system may bear no semblance to random sampling or a  thermodynamic limit. We develop here an iterative parameter-free model estimator using only the mathematical formalism of statistical physics to define a free energy of data, and show that minimizing this free energy corresponds to linear and higher-order data regressions. 
Over-fitting is a major problem in the analysis of under-determined systems.
By decoupling an iterative Rao-Blackwell estimator update step from an update--consistent stopping criterion, we demonstrate that our Free Energy Minimization (FEM) approach infers coupling strengths in non-equilibrium kinetic Ising models, outperforming previous approaches particularly in the large coupling variability and small sample size regimes. Real data is always a stringent test of model inference so we demonstrate applications of FEM to infer biological and financial networks from neuronal activities and currency fluctuations.

\section*{Iterative stochastic causality inference from Free Energy Minimization}
The elegant mathematical formalism developed by Schwinger provides a natural connection between expectation values $m=\langle \sigma \rangle$ of microstates $\sigma$ and expectation values $\langle E_i \rangle_m$ of observables $E_i$ conditioned on $m$~\cite{schwinger1953,toms2007schwinger}. 
We will use it to implement a Rao-Blackwell estimator update. As a concrete illustration, let us start with a kinetic Ising model in which a vector $\sigma$ of $N$ spins $\sigma_i(t)=\pm1$ is stochastically updated based on the following conditional probability
\begin{equation}
\label{eq:kIsingProb}
P(\sigma_i(t+1)=\pm1|\sigma(t)) = \frac{\exp (\pm H_i(\sigma(t)))}{\exp(H_i(\sigma(t))) + \exp(-H_i(\sigma(t)))}
\end{equation}
with a local field $H_i(\sigma(t)) \equiv \sum_j W_{ij} \sigma_j(t) $. 
Our goal is to infer the coupling strength $W_{ij}$ that minimizes the discrepancy between observed $\sigma_i(t+1)$ and model expectation ${\langle \langle \sigma_i(t+1) \rangle \rangle_{\sigma(t)}}\equiv \sum_{\rho=\pm1} \rho P(\sigma_i(t+1)=\rho|\sigma(t)).$ For the kinetic Ising model, ${\langle \langle \sigma_i(t+1) \rangle \rangle_{\sigma(t)}}=\tanh H_i(\sigma(t)).$ 



To implement a Rao-Blackwell scheme of estimator improvement $H_i^{\textrm{new}}(m) \leftarrow \langle E_i\rangle_m,$ we first define a moment generating function, $Z(J,\beta)=\sum_{t} \exp( J\cdot \sigma(t) - \beta E_i(t) )$, which is 
a function of a vector parameter $J$, a scalar parameter $\beta,$ and a `data energy' $E_i(t)$ that we will define below. 
A convex free energy $F=\log Z$ can be used  to obtain expectation values of spin activities by differentiation,
\begin{equation}
\label{eq:hbeta}
\frac{\partial F}{\partial J_i} = \frac{\sum_{t} \sigma_i(t) \exp ( J\cdot \sigma(t) - \beta E_i(t) )}{\sum_{t} \exp (J\cdot \sigma(t) - \beta E_i(t) )} = \langle \sigma_i \rangle_J \equiv m_i(J).
\end{equation}
As usual, a convex dual free energy $G$ can be defined to make the expected activity vector $m$ the independent variable, and $J(m)$ the dependent vector, by using the convexity preserving Legendre transform $F(J)+G(m)=J\cdot m.$ The expectation value of $E_i$ is obtained by differentiation (identifying $\langle E_i \rangle_{J(m)} \equiv \langle E_i \rangle_{m}$),
\begin{equation}
\label{eq:dG}
\frac{\partial G}{\partial \beta}=-\frac{\partial F}{\partial \beta} = \frac{\sum_{t} E_{i}(t) \exp ( J\cdot \sigma(t) - \beta E_i(t) )}{\sum_{t} \exp (J\cdot \sigma(t) - \beta E_i(t) )} = \langle E_i  \rangle_{m}.
\end{equation}
The free energy $G(m, \beta) = \beta \langle E_i  \rangle_{m} - S$  
where $S$ 
is the Shannon entropy of data. At $\beta=0,$ minimizing the free energy is exactly maximizing the entropy, making every sample equally valuable. 
At its minimum, $m^*,$ we have $J(m^*)=\partial_mG(m^*)=0,$ and this is the value of $J$ about which we will expand, hence the term Free Energy Minimization (FEM).

We now turn to finding an appropriate $E_i.$ Consider
\begin{equation}
\label{eq:E}
E_i(t) \equiv \frac{\sigma_i(t+1)}{\langle \langle \sigma_i(t+1) \rangle \rangle_{\sigma(t)}} H_i(\sigma(t)),
\end{equation}
and  define the Rao-Blackwell conditional expectation update: $H_i(m)^{\textrm{new}} \leftarrow \langle E_i \rangle_{m}.$ Intuitively, if the observation $\sigma_i(t+1)$ is larger/smaller than the corresponding model expectation $\langle \langle \sigma_i(t+1)\rangle \rangle_{\sigma(t)},$ this update increases/decreases $H_i(\sigma(t))$ proportionally to the discrepancy ratio between the observation and the model expectation, including the  sign. The differential geometry of $G(m, \beta)$ around its minimum $m^*$ then gives
$W_{ij}^{\textrm{new}} = \sum_k \langle \delta E_i \delta \sigma_k  \rangle_{m^*} [C^{-1}]_{kj}$ as a matrix multiplication, where $\delta f \equiv f -\langle f\rangle_{m^*}$ and $C_{jk} \equiv \langle \delta\sigma_j\delta\sigma_k\rangle_{m^*}$ (see SI Text 1 for the detailed derivation).

The second crucial aspect for small sample size inference is to find a suitable stopping criterion for the Rao-Blackwell update. 
We consider the overall discrepancy between ${\sigma_i(t+1)}$ and ${\langle\langle \sigma_i(t+1) \rangle\rangle_{\sigma(t)}}$:
\begin{equation}
\label{eq:D}
D_i(W)\equiv\sum_{t} \big[ \sigma_i(t+1) - \langle \langle \sigma_i(t+1) \rangle \rangle_{\sigma(t)} \big]^2.
\end{equation}
The minimum of $D_i(W)$ is the closest we can approach a fixed point of the update iteration, consistent with Eq.~(\ref{eq:E}) and the Rao-Blackwell expectation. Therefore, we stop the iteration when $D_i(W)$ starts to increase.

To summarize inference with FEM: 
(i) Compute $H_i(\sigma(t)) \equiv \sum_j W_{ij} \sigma_j (t)$ (initialize with a random $W_{ij}$);
(ii) Compute $E_i(t)$ as defined in Eq.~(\ref{eq:E});
(iii) Extract $W_{ij}^{\textrm{new}}= \sum_k \langle \delta E_i \delta \sigma_k  \rangle_{m^*} [C^{-1}]_{kj};$ 
(iv) Repeat (i)-(iii) until $D_i(W)$ starts to increase;
(v) Compute (i)-(iv) in parallel for every index $i \in \{1, 2, \cdots, N\}$.

\begin{figure*}
\centering
\includegraphics[width=16cm]{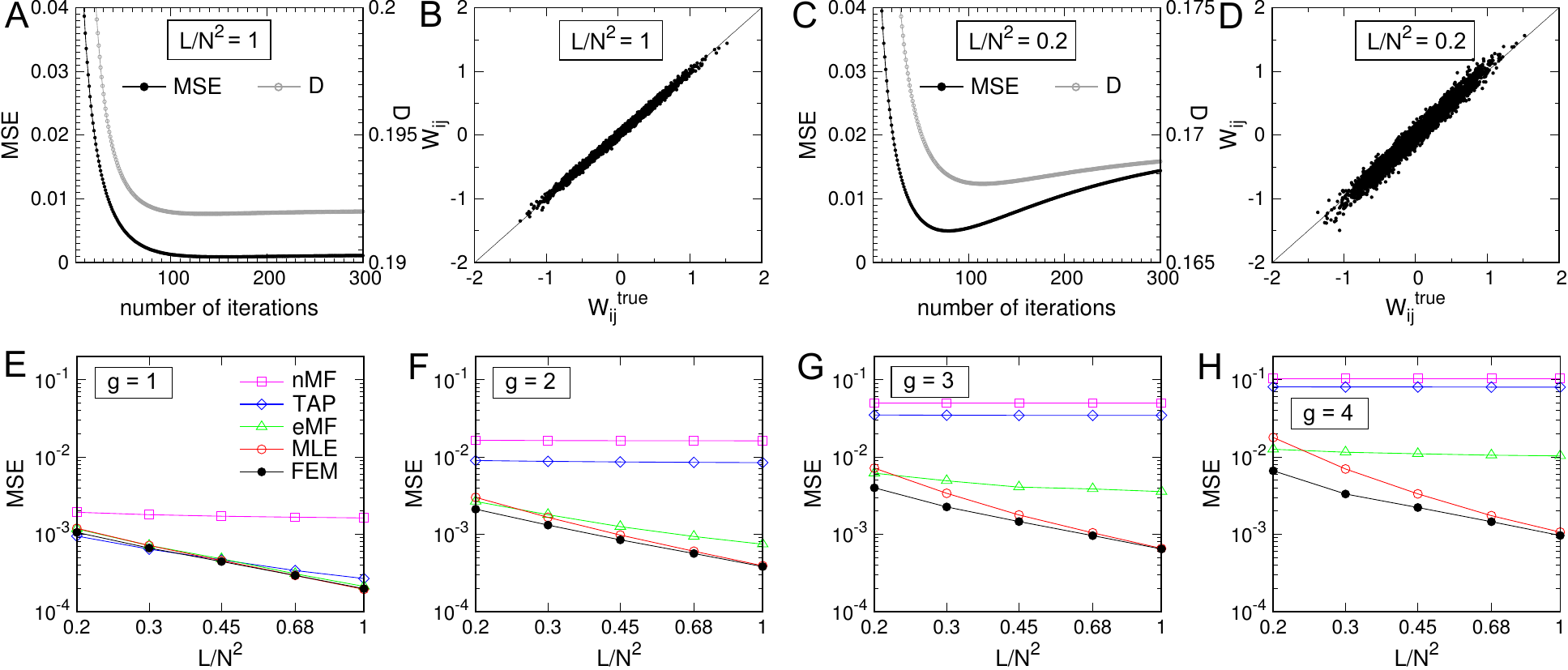}
\caption{ \label{fig:fig1}Network inference for the kinetic Ising model. Inference error (MSE, black) and discrepancy ($D$, gray) are shown as function of number of iterations for large observed configurations, $L/N^2=1$ (A) and few observed configurations, $L/N^2=0.2$ (C). Predicted couplings versus actual couplings for $L/N^2=1$ (B) and $L/N^2=0.2$ (D). The inference errors are obtained for na\"ive mean-field (nMF), Thouless-Anderson-Palmer (TAP), exact mean-field (eMF), Maximum Likelihood Estimation (MLE), and Free Energy Minimization (FEM), for various number of observed configurations, $L/N^2$ from $0.2$ to $1$ in the limit of weak coupling, $g=1$ (E), and in the limit of stronger coupling, $g=2$ (F), $g=3$ (G), and $g=4$ (H). A system size $N=100$ is used. A learning rate $\alpha=1$ is used for MLE.
}
\end{figure*}

\section*{Results}
\subsection*{Kinetic Ising model}
We first tested FEM on the inference of connection weights $W_{ij}$ ($\neq W_{ji}$) in the kinetic Ising model, which is often used as a benchmark for stochastic causality inference.
The Sherrington-Kirkpatrick (SK) model assumes $W_{ij}$ are normally distributed with zero mean and variance equal to $g^2/N$~\cite{Sherrington1975}.
In the limit of large sample size (large $L/N^2$), our iterative method decreases the mean square error, 
MSE = $N^{-2} \sum_{i,j=1}^N (W_{ij} - W_{ij}^{\textrm{true}})^2$, as the number of iterations increases (Fig.~\ref{fig:fig1}A).
We obtain good agreement between true and predicted weights (Fig.~\ref{fig:fig1}B).
In real world problems, $W_{ij}^{\textrm{true}}$ is inaccessible so MSE cannot be defined.
However, $D_i(W)$ in Eq.~(\ref{eq:D}) is an alternative measure of the discrepancy between observation $\sigma_i(t+1)$ and model expectation.
The discrepancy measures $D_i(W)$ are independent for each spin $i.$ 
We checked that MSE and $D=N^{-1}\sum_{i=1}^N D_i(W)$ change similarly during iterations.
More importantly, for small sample sizes (small $L/N^2$), MSE and $D$ decrease  with iterations initially, 
but start to increase after some number of iterations (Fig.~\ref{fig:fig1}C). 
For the kinetic Ising model, $D_i(W)=4\sum_{t} [1-P(\sigma_i(t+1)|\sigma(t))]^2$ 
with the transition probability, $P(\sigma_i(t+1)|\sigma(t))$ in Eq.~(\ref{eq:kIsingProb}).
Therefore, decreasing $D_i(W)$ can only result from $P(\sigma_i(t+1)|\sigma(t))$ saturating the causal relation between observations, $\sigma(t)$ and $\sigma_i(t+1)$, through $W.$ Distinct spins indexed by  $i$ often require different numbers of iterations.
Stopping the iteration for spin $i$ when $D_i(W)$ saturates leads to accurate inference with minimal computation.
For limited data (e.g. $L/N^2 = 0.2$), these stopping criteria lead to accurate inference (Fig.~\ref{fig:fig1}D) without over-fitting.

Now we compare the inference performance of our method with other representative methods~\cite{Roudi2011, Mezard2011, Zeng2013}: na\"ive mean field (nMF), Thouless-Anderson-Palmer mean field (TAP), exact mean field (eMF), and maximum likelihood estimation (MLE). 
MLE requires maximizing the data likelihood, ${\cal P} = \prod_{t=1}^{L-1} \prod_{i=1}^N P(\sigma_i(t+1)|\sigma(t))$, and uses gradient ascent to update $W_{ij}$ incrementally through $W^{\textrm{new}}_{ij} = W_{ij} + {\alpha}/({L-1}) {\partial \log {\cal P}}/{\partial W_{ij}}$~\cite{Roudi2011, Zeng2013},
where the learning rate $\alpha$ is an undetermined parameter controlling the updating speed.
In contrast, the maximizing condition ($\partial \log {\cal P}/{\partial W_{ij}}=0$) and mean-field approximations provide matrix equations,
$W=A^{-1}BC^{-1}$, where matrices $B_{ij} = \langle \delta \sigma_i(t+1) \delta \sigma_j(t) \rangle$ and $C_{ij}= \langle \delta \sigma_i(t) \delta \sigma_j(t) \rangle$ represent time-delayed and equal-time correlations in data, and $A$ are diagonal matrices, which are different for nMF, TAP, and eMF (SI Text 2 has brief reviews of these mean-field methods).

For weak coupling ($g=1$), TAP, eMF, MLE and FEM have similar inference accuracy that increases with sample size (Fig.~\ref{fig:fig1}E).
nMF showed poor accuracy independent of data size, since the zeroth-order mean-field approximation works only for very weak coupling strengths~\cite{Roudi2011}.
As we further increase coupling strength, the other two mean-field methods, TAP and eMF also start to give less accurate results than MLE and FEM (Fig.~\ref{fig:fig1}F-H). 
For large sample size ($L/N^2>1$), our iterative method, FEM, works as well as standard MLE. For small sample size, however, FEM provides better accuracy than MLE.
For example, the inference error (MSE) of FEM is approximately 4 times lower than that of MLE for $L/N^2=0.2$ and $g=4.$
In addition to inference accuracy, FEM has two advantages in computation. 
First, the FEM update is multiplicative and not incremental, while MLE updates (using conjugate gradient ascent or some other numerical maximization) have an undetermined parameter, the learning rate $\alpha,$ which needs to be determined.
A very large rate ($\alpha=3$) leads to loss of convergence, whereas a very small rate ($\alpha=0.5$) leads to many iterations with infinitesimal updates. We set $\alpha=1$.
Second, FEM requires 20 times fewer updates than MLE (Fig.~\ref{fig:figS1}A), which reduces computation time a 100-fold (Fig.~\ref{fig:figS1}B).

\begin{figure*}
\centering
\includegraphics[width=16cm]{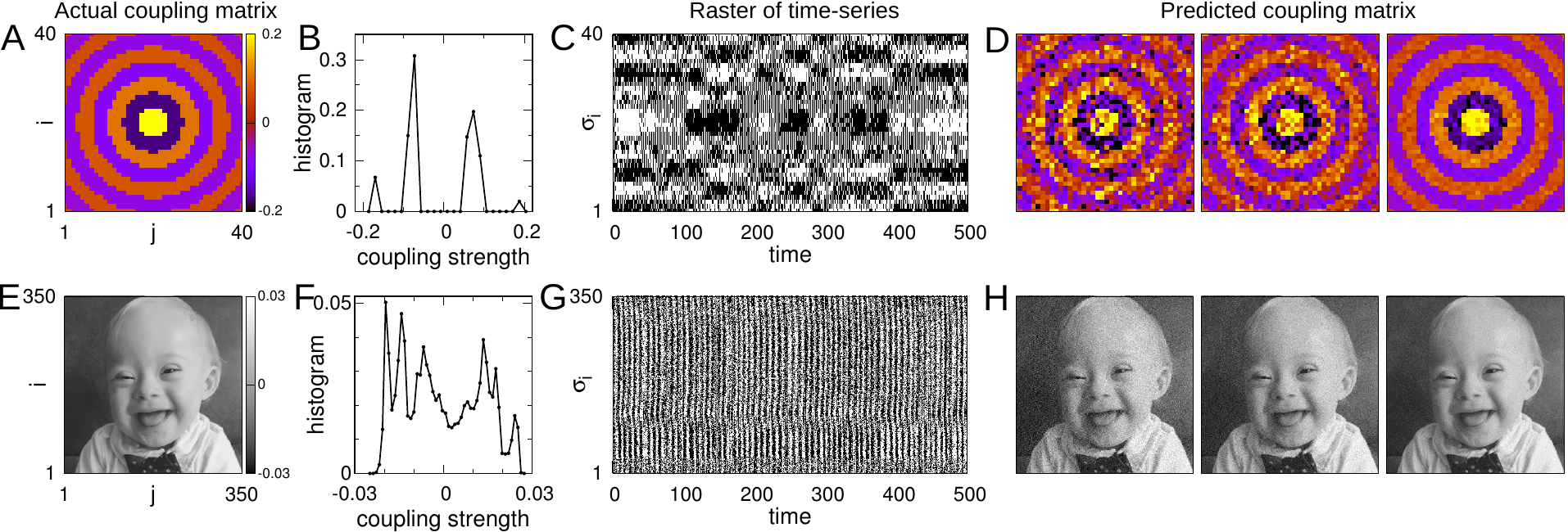}
\caption{ \label{fig:fig2}Effectiveness of FEM in inferring network with specific structures. Given true coupling weights of $N=40$ (A) and  $350$ (E) spin variables with non-Gaussian distributions, typical time-series of their activities are generated (C, G). Predicted coupling weights are obtained for different data lengths $L/N^2=0.5$, $1$, and $4$ from left to right (D, H). The image is converted from the 2018 Gerber baby's photograph (with permission from Gerber).
}
\end{figure*}

\begin{figure}
\centering
\includegraphics[width=8.cm]{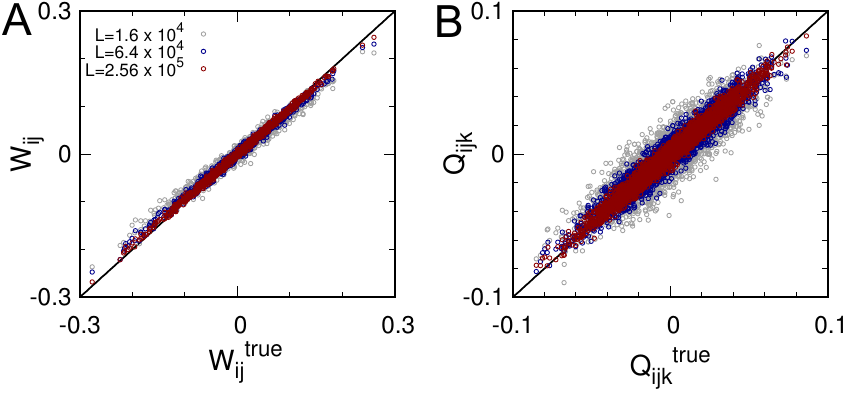}
\caption{ \label{fig:fig3}Accurate inference of higher-order coupling strengths.
Linear (A) and quadratic (B) coupling strengths in the nonlinear kinetic Ising model 
are predicted from FEM. Here the true coupling strengths are normally distributed with a system size $N=40$.
Three different data lengths, $L=1.6\times10^4$ (gray), $6.4\times10^4$ (blue) and $2.56\times10^5$ (red), are examined. 
}
\end{figure}

To further demonstrate the effectiveness of FEM, we show two examples of inferred networks when $W_{ij}$ has more general coupling distributions than the SK model, as real systems often deviate strongly from normally-distributed coupling strengths.
In the first example, the spins have alternating bands of positive and negative couplings modulated by distance as $|W_{ij}| = W_{0}/\log(R_{ij})$, where $R_{ij}$ represents the radius of the circle (Fig.~\ref{fig:fig2}A). The couplings are non-normally distributed (Fig.~\ref{fig:fig2}B). The spin raster scan exhibits nontrivial structure (Fig.~\ref{fig:fig2}C), reminiscent of binocular rivalry~\cite{Moreno-Bote:2007a}. 
As the number of observed configurations increases, the predicted coupling strengths (Fig.~\ref{fig:fig2}D) approach their true values (Fig.~\ref{fig:fig2}A).
In the second, the 2018 Gerber baby's photograph was used as the heatmap of the coupling matrix (Fig.~\ref{fig:fig2}E). These couplings are also non-normally distributed (Fig.~\ref{fig:fig2}F) with periodic bursting in the simulated spin raster scan (Fig.~\ref{fig:fig2}G), but the couplings are still predicted well (Fig.~\ref{fig:fig2}H).

Our formulation, based on the differential geometry of the data free energy, automatically includes higher-order regression equations for the local field $H_i (\sigma)$ (SI Text 1).
For example, we checked higher-order inference with FEM by using a generalized kinetic Ising model with linear and quadratic couplings, $H_i(\sigma(t))=\sum_{j} W_{ij}\sigma_{j}(t) + \sum_{j,k} Q_{ijk} \sigma_j(t)\sigma_k(t)/2$, where $W_{ij}$ and $Q_{ijk}$ are normally distributed. The quadratic couplings are symmetric ($Q_{ijk} = Q_{ikj}$) and have no self-interactions ($Q_{ijj}=0$) since $\sigma_j^2 = 1.$ The number of $Q_{ijk}$ parameters is $N^2(N-1)/2.$ 
The recovery of both linear and quadratic couplings is evident (Fig.~\ref{fig:fig3}).

\begin{figure*}
\centering
\includegraphics[width=15cm]{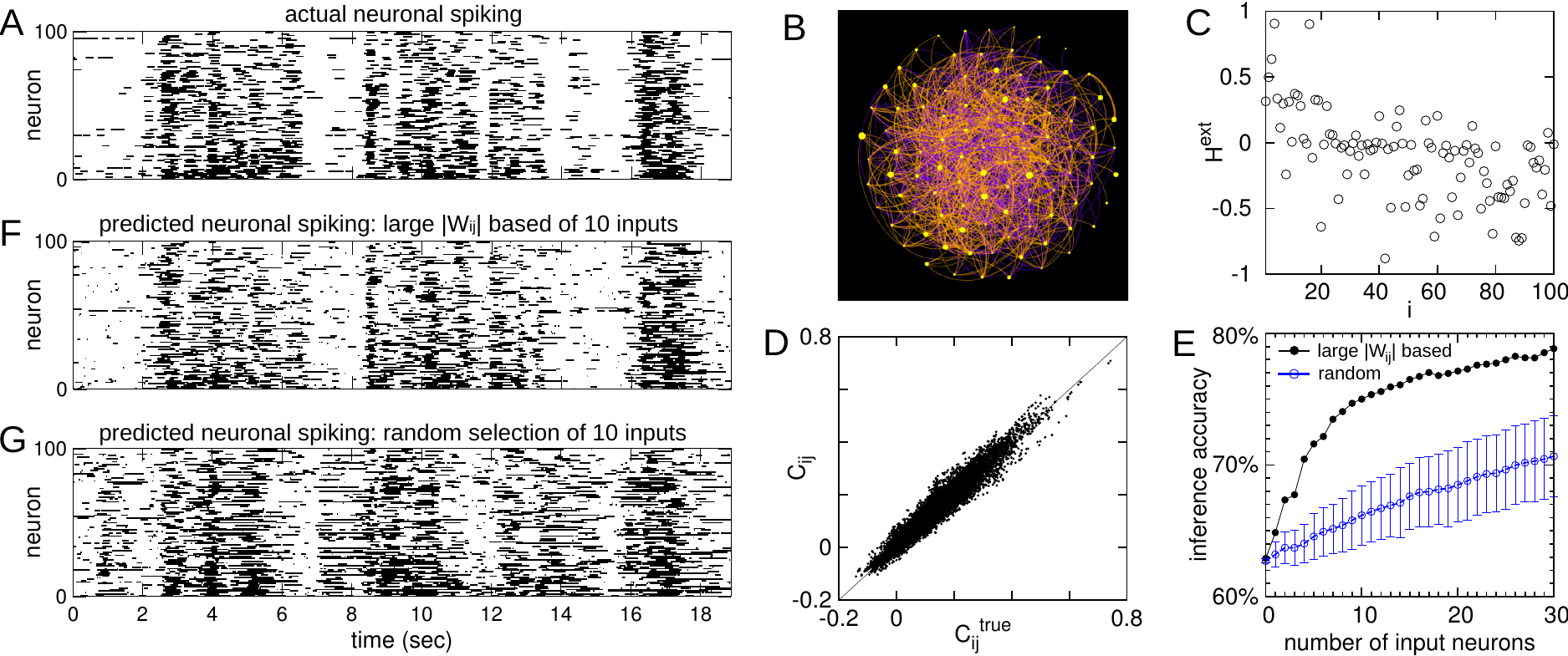}
\caption{\label{fig:fig4} Inference of coupling strengths between neurons, external local fields and neuronal activities. From activities of 100 neurons (A), neuronal network (B) and external local field $H_i^{\textrm{ext}}$ (C) are predicted. The red and blue edges represent positive and negative couplings, respectively. Edge direction is clock-wise. Inferred correlation covariances $C_{ij}$ are compared with actual correlation covariances $C_{ij}^{\text{true}}$ (D). Inference accuracy of remaining neuronal activities versus number of input neurons selected based on large $|W_{ij}|$ (filled black circles), and randomly selected (empty blue circles). Error bars represent the standard deviation from 50 random trials (E). Neuronal activities are reconstructed with 10 input neurons, selected based on large $|W_{ij}|$ (F), and randomly selected (G).}
\end{figure*}

\subsection*{Neuronal network}
We applied our method to infer a neuronal network from temporal neuronal activities in the tiger salamander ({\it Ambystoma tigrinum}) retina~\cite{Marre2017}.
The multi-channel experiment recorded stochastic firing patterns of 160 neurons when the salamander retina was stimulated by a film clip of fish swimming. 
As in Ref.~\cite{Tkacik2014}, 
we considered only the 100 most active neurons. 
After processing the data (SI Text 3; Fig.~\ref{fig:fig4}A), we inferred the neuronal network governing the local field, $H_i(\sigma(t))=H_{i}^{\textrm{ext}} + \sum_{j} W_{ij}\sigma_{j}(t)$.
Here we included a constant bias external field $H_{i}^{\textrm{ext}}$ for neuron $i$ to consider the persistent silence of neurons.
We inferred the neuronal network weights $W_{ij}$ (Fig.~\ref{fig:fig4}B), and the external local fields for each neuron by using $H_{i}^{\textrm{ext}} = \langle H_i \rangle - \sum_{j} W_{ij} \langle \sigma_{j} \rangle $.  
The external local fields are mostly negative, which implies that neuronal activities are biased to be silent (Fig.~\ref{fig:fig4}C).

The true couplings are unknown for this system. As a  validation, with the $H_{i}^{{\textrm{ext}}}$ and $W_{ij}$ we determined, we simulated neuronal activities. 
We found agreement between the covariances of neuronal activities $C_{ij}= \langle \delta \sigma_i(t) \delta \sigma_j(t) \rangle$ 
of the observed and simulated data (Fig.~\ref{fig:fig4}D).
For a more stringent validation, we reconstructed the full neuronal activities from specific `pinned' neuron activities, representing inputs.
Fixing the time sequences $\sigma_j(t)$ of specific chosen input neurons $j \in I$, we reconstructed the activities $\sigma_i(t+1)$ of the remaining neurons $i \not\in I$. 
As a control, we selected the input neurons at random and compared them with input neurons selected on the basis of the coupling strength $|W_{ij}|$ as the input set $I.$
As more input neurons are considered, the reconstruction predicts $\sigma_i(t+1)$ more accurately (Figs.~\ref{fig:fig4}E and~\ref{fig:figS3}). 
 Pinning the activities of only $|I|=10$ strongly coupled neurons gave predicted activities of the remaining 90 neurons that were very close to the observed activities (Fig.~\ref{fig:fig4}F), in contrast to predicted activities obtained by pinning randomly selected sets of 10 input neurons (Fig.~\ref{fig:fig4}G).

\subsection*{Currency network}

\begin{figure*}
\centering
\includegraphics[width=15cm]{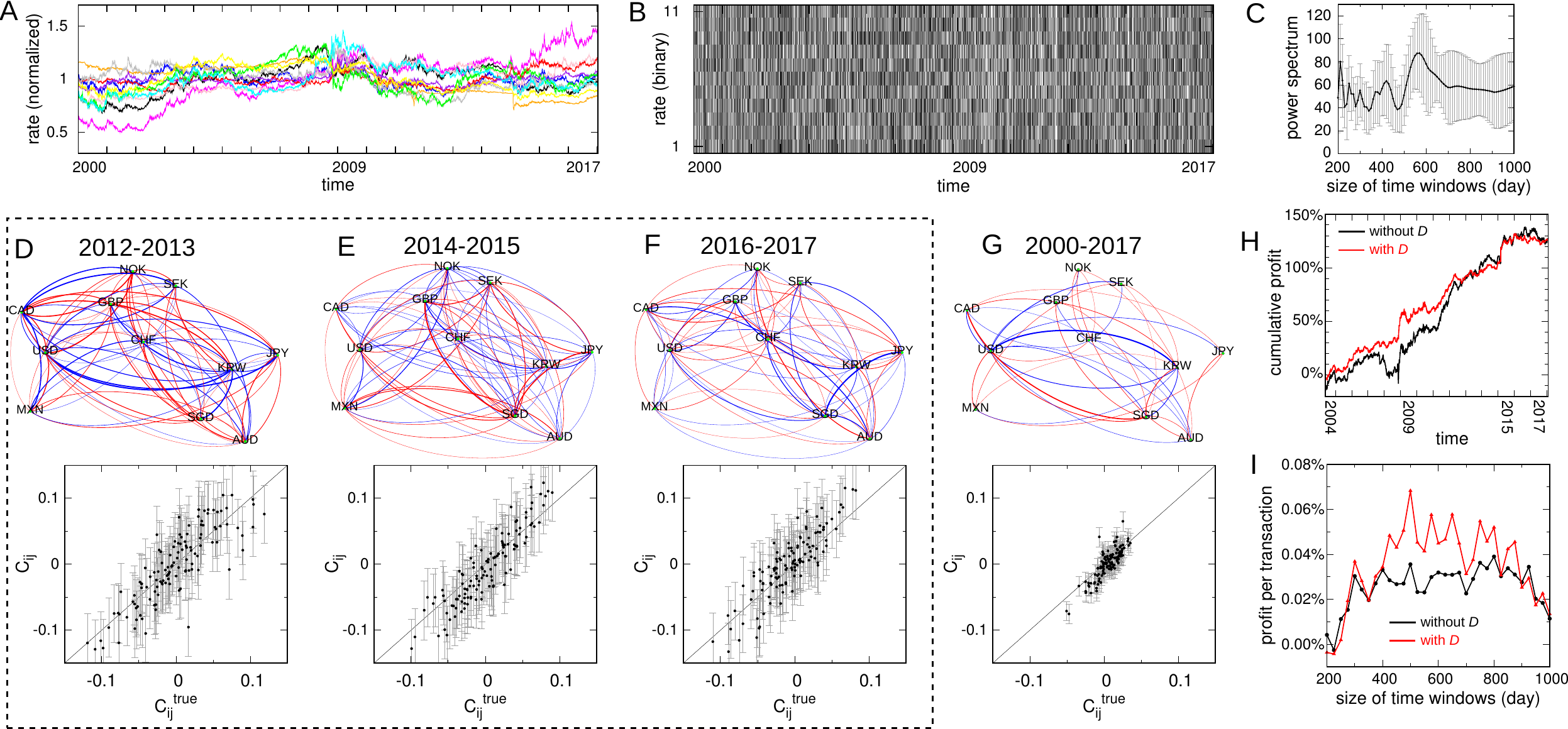}
\caption{ \label{fig:fig5}Inference of coupling strengths between currency exchange rates. Normalized exchange rates relative to EUR of 11 currencies are plotted with different colors representing distinct currencies (A). A raster representation of binarized exchange rate fluctuations is plotted with black dots representing increase, white dots  decrease. Average power spectrum obtained from a Fourier transform of exchange rate fluctuations versus time-window size in which error bar represents standard deviation from different currencies (C). The currency networks are predicted for different periods, e.g. from the years of 2012 to 2013 (D), 2014 to 2015 (E), and 2016 to 2017 (F). The network for the whole data, from 2000 to 2017, is also predicted (G). The red and blue edges represent positive and negative couplings, respectively. Edge direction is clock-wise. Predicted covariances are shown to compare with observed covariances $C_{ij}^{\text{true}}$ (D-G, lower). Cumulative profit versus time period for various time-window sizes (H). Profit per transaction using our strategy is plotted as a function of time-window size (I).  
}
\end{figure*}

Finally, we apply our method to another difficult and representative stochastic problem, currency exchange rate fluctuations.
We obtained time series of currency exchange rates from January 2000 to December 2017~\cite{BankItaly}, and examined exchange rates denominated in Euro (EUR) 
of $11$ actively traded currencies (Fig.~\ref{fig:fig5}A). 
First, we concentrate on the daily fluctuations of the exchange rates, since most financial analyses center on price increments rather than absolute prices~\cite{Pincus2004}.
We binarize the real-valued rates to concentrate on the sign of their daily fluctuations (Fig.~\ref{fig:fig5}B).
We defined the binarized rate $\sigma_i(t)=1$ for a day-to-day increase of exchange rate $i$ at time $t$ ($r_i(t) > r_i(t-1))$,
and $\sigma_i(t)=-1$ for the decrease. If there was no change ($r_i(t) = r_i(t-1)$), we set $\sigma_i(t)=\sigma_i(t-1)$.
Second, we divide the data for different periods to investigate the time dependence of the couplings between exchange rates.
Using the Fourier transform of the binarized time series, we identified a characteristic period, 550 business days ($\sim$ 2 years), of the fluctuations (Fig.~\ref{fig:fig5}C). 
We inferred the currency network weights $W_{ij}$ separately in two year periods, shown here (Figs.~\ref{fig:fig5}D-F, upper) for the three periods  2012-2013, 2014-2015, and 2016-2017.
We found agreement between the covariance $C_{ij}=\langle \delta \sigma_i(t) \delta \sigma_j(t) \rangle$ of the observed currency data and that of the simulated currency data using $H_i(\sigma(t))=H_{i}^{\textrm{ext}} + \sum_{j} W_{ij}\sigma_{j}(t)$ (Figs.~\ref{fig:fig5}D-F, lower).
In contrast, when we estimated the currency network using the data for the entire period 2000-2017, the network had weaker connections and smaller covariances $C_{ij}$ compared to the time-dependent analysis (Figs.~\ref{fig:fig5}G)

The raw exchange rate data is continuous. Is our binarized inference of any practical value? To address this, we simulated a currency trade strategy, and checked if the strategy was profitable. Using only data within a time window of a period $T$,  $\{\sigma(t-T+1), \sigma(t-T+2), \cdots, \sigma(t) \},$
we predicted the currency fluctuations $\sigma(t+1)$ on the next day.
For the trade simulation, we considered a hedging trader who buys one currency with 1 EUR and sells one currency with 1 EUR.
To earn profits, the trader is supposed to sell/buy a currency that has the highest probability of increase/decrease in exchange rate:
the currency $sell = \arg\max_i P(\sigma_i(t+1)=+1|\sigma(t))$  and the currency $buy = \arg\max_i P(\sigma_i(t+1)=-1|\sigma(t))$.
Then, a daily profit can be defined as ${\text{profit} (t)} = r_{sell}(t+1)/r_{sell}(t) - r_{buy}(t+1)/r_{buy}(t)$.
We calculated cumulative profits of the trade simulation from 2004 to 2017 with various time window sizes that we considered as past information (Fig.~\ref{fig:fig5}H for $T = 500$ days). 
Hedging strategies profit from market volatility and, indeed, our trade simulation showed large profits when the exchange rates had large fluctuations (Fig.~\ref{fig:fig5}A).
The window size $T$ had an optimal period of 500-750 business days (Fig.~\ref{fig:fig5}I).
For a more refined strategy, we considered the quality or accuracy of our inference by probing the discrepancy $D_i(W)$ in Eq.~(\ref{eq:D}).
Instead of trading every day, we traded only on the days when the discrepancy at that day, 
$D(t) \equiv \sum_{i} \big[ \sigma_i(t) - \langle \langle \sigma_{i}(t) \rangle \rangle _{\sigma(t-1)} \big]^2$, was lower than the average $T^{-1} \sum_{t=1}^T D(t)$ for a fixed window size $T$. 
This strategy doubled the profits per transaction (Figs.~\ref{fig:fig5}H and \ref{fig:fig5}I), showing that the discrepancy $D_i(W)$ is a useful measure of model accuracy.

\section*{Discussion}
We demonstrated that under-determined stochastic systems can be inferred in a conceptually simple and computationally efficient manner using the mathematical framework of statistical physics.
Since network inference is an important subject, many different approaches have been developed.
Equilibrium approaches assume symmetric interactions ($W_{ij}=W_{ji}$) between node $i$ and node $j$, and estimate the pair-wise interaction strengths that can maximally explain the observed static patterns of network activity in brains~\cite{Tkacik2014, Tkacik2015, Watanabe2013}, proteins~\cite{Mora2010, Weigt2009}, and stock markets~\cite{Bury2012}.
In contrast, non-equilibrium approaches do not assume symmetry, and infer asymmetric causal relations between nodes that can better explain  dynamic patterns of network activity~\cite{Zeng2013}.
Causality inference for non-equilibrium models (e.g., using recurrent neuronal networks) is computationally expensive.
Although mean-field methods have been introduced to circumvent this practical problem~\cite{Roudi2011}, these approximation methods only work for weak-interaction regimes with large sample size.
All small sample size inference must contend with over-fitting so the key feature of our approach was to consistently decouple the model update step and a discrepancy measure that is similar to Expectation Maximization. This decoupling allowed us to iterate with a multiplicative model update, and to stop when the discrepancy measure quantifies that the multiplicative update has saturated. 
We derived this within a standard statistical physics formulation~\cite{schwinger1953,toms2007schwinger},
so no ad hoc averaging or approximation steps were involved. We demonstrated that our method outperfoms others in inferring the asymmetric interactions of the kinetic Ising model, especially in strong-interaction regimes, and particularly when available data was limited. 
Another aspect of small sample size inference is that longer time-scale modulation of couplings can be uncovered. This is of considerable practical import as we demonstrated with the currency exchange rate network.

FEM has several computational merits. Besides having no incremental learning rate that requires tuning, 
the method is parallelizable and scalable:
We computed results for the kinetic Ising model with up to $N=5000$ interacting spins, determining $2.5 \times 10^7$ parameters (Fig.~\ref{fig:figS4}).
We also demonstrated that the method can infer not only linear interactions but also higher-order interactions.
Moreover, FEM is generalizable to systems with any number of discrete states, although we focused on binary stochastic systems here.
Uncovering hidden nodes for stochastic network inference~\cite{Hoang2018} is an exciting avenue for future work.


\acknow{Ga\v{s}per Tka\v{c}ik  generously provided the neuronal activity data. We thank Changbong Hyeon and Arthur Sherman for comments on the manuscript.
This work was supported by Intramural Research Program of the National Institutes of Health, NIDDK (D.-T.H.,V.P.), and by Basic Science Research Program through the National Research Foundation of Korea (NRF) funded by the Ministry of Education (2016R1D1A1B03932264) and the Max Planck Society, Gyeongsangbuk-Do and Pohang City (J.J.).}

\showacknow 


\bibliography{pnas-network}

\clearpage
\setcounter{equation}{0}
\setcounter{figure}{0}
\setcounter{table}{0}
\setcounter{page}{1}
\makeatletter
\renewcommand{\theequation}{S\arabic{equation}}
\renewcommand{\thefigure}{S\arabic{figure}}

\section*{Supporting Information (SI)}
\subsection*{SI Text 1: Schwinger's source formalism}
Here, we derive the differential geometry of $\langle E_i \rangle$ in terms of $\langle \sigma \rangle$ dependency by using  Schwinger's source formalism~\cite{schwinger1953,toms2007schwinger}.
This is a model-free approach, because we do not assume a specific functional form of $\langle E_i \rangle$ at the beginning.
First, we defined the moment generating function,
\begin{equation}
\label{eq:Z}
Z(J, \beta) = \sum_t \exp (J \cdot \sigma(t) - \beta E_i(t)).
\end{equation}
The log partition function, $F=\log Z$, allows the computation of expectation values of $\sigma$ and $E_i$ simply by differentiation
\begin{align}
\label{mJ}
\frac{\partial F}{\partial J} &= \frac{\sum_{t} \sigma(t) \exp(J \cdot \sigma(t) - \beta E_i(t))}{Z}= \langle \sigma \rangle_{J, \beta} = m(J),\\
- \frac{\partial F}{\partial \beta} &= \frac{\sum_{t} E_i(t) \exp(J \cdot \sigma(t) - \beta E_i(t))}{Z}= \langle E_i \rangle_{J, \beta}.
\end{align}
Here, the activity expectation $m(J)$ depends on $J$.
We can make the observable expectation $m$ the independent variable, and the control parameter $J$ the dependent variable by using a Legendre transform:
\begin{equation}
\label{legendre}
F(J, \beta)+G(m, \beta)=J\cdot m.
\end{equation}
Defining a normalized probability,
\begin{equation}
P(\sigma(t)) \equiv \exp(J\cdot \sigma(t) - \beta E_i(\sigma(t)) +F)
\end{equation} 
in Eq.~(\ref{eq:Z}), it is straightforward to show that
\begin{equation}
G(m, \beta) = \beta \langle E_i \rangle_{J, \beta} - S,
\end{equation}
with the Shannon entropy appearing naturally,
\begin{equation}
S = - \sum_{t} P(\sigma(t)) \log P(\sigma(t)).
\end{equation}
Then, the duality between the free energies $F$ and $G$ through their Legendre transform in Eq.~(\ref{legendre}) leads to
\begin{eqnarray}
\label{eq:J}
&& \frac{\partial G}{\partial m} = J,\\
&& \frac{\partial G}{\partial \beta} = -\frac{\partial F}{\partial \beta} = \langle E_i \rangle_{J, \beta}.
\end{eqnarray}
Therefore, once we know the free energy $G(m, \beta)$, it is straightforward to obtain $\langle E \rangle_{J, \beta}$.
For our purposes, however, it is unnecessary to obtain $G(m, \beta)$ for all values of $m,$ 
as it suffices to know the function at minimum, because the free energy is minimized at the data expectation:
$m^*=\langle \sigma \rangle_{J=0, \beta=0}$.
Note that $J=0$ imposes the minimum condition ($\partial G / \partial m = 0$) in Eq.~(\ref{eq:J}).
Then, we have the Taylor expansion of  $G(m, \beta)$ at $m=m^*:$
\begin{align}
G(m,\beta) &= G(m^*,\beta) + \frac{1}{2} \sum_{j,k} \bigg[ \frac{\partial^2 G}{\partial m_j \partial m_k} \bigg]^* (m_j-m_j^*)(m_k-m_k^*) \nonumber \\
&+ \frac{1}{6} \sum_{j,k,l} \bigg[ \frac{\partial^3 G}{\partial m_j \partial m_k \partial m_l} \bigg]^* (m_j-m_j^*)(m_k-m_k^*)(m_l-m_l^*)\nonumber \\
& +{\cal O}(\delta^4 m) 
\end{align}
where the derivatives $[\cdot]^*$ are taken at $m=m^*$.
Differentiating the expanded $G(m, \beta)$ with respect to $\beta$ leads to
\begin{align}
\label{eq:dGdb}
\frac{\partial G(m,\beta)}{\partial \beta} &= \frac{\partial G(m^*,\beta)}{\partial \beta} - \sum_{j,k} \frac{\partial m_k^* }{\partial \beta} \bigg[ \frac{\partial^2 G}{\partial m_j \partial m_k} \bigg]^* (m_j-m_j^*) \nonumber \\
&+ \frac{1}{2} \sum_{j,k} \frac{\partial }{\partial \beta} \bigg[ \frac{\partial^2 G}{\partial m_j \partial m_k} \bigg]^* (m_j-m_j^*)(m_k-m_k^*)\nonumber \\
&- \frac{1}{2} \sum_{j,k,l} \frac{\partial m_l^* }{\partial \beta} \bigg[ \frac{\partial^3 G}{\partial m_j \partial m_k \partial m_l} \bigg]^* (m_j-m_j^*)(m_k-m_k^*) \nonumber \\
&+{\cal O}(\delta^3 m) .
\end{align}
Now, we calculate each derivative in Eq.~(\ref{eq:dGdb}):
\begin{itemize}
\item[(i)]
\begin{equation}
-\frac{\partial m_k}{\partial \beta} = \frac{\partial}{\partial \beta} \bigg[ \frac{\sum_{t} \sigma_k(t) \exp (J\cdot \sigma(t) - \beta E_i(t))}{\sum_{t} \exp (J\cdot \sigma(t) - \beta E_i(t))} \bigg] =\langle \delta E_i \delta \sigma_k \rangle.
\end{equation}
\item[(ii)]
\begin{equation}
\frac{\partial^2 G}{\partial m_j \partial m_k} = \frac{\partial J_k}{\partial m_j}=[C^{-1}]_{jk},
\end{equation}
where 
\begin{align}
C_{jk} &= \frac{\partial m_j}{\partial J_k}= \frac{\partial}{\partial J_k} \bigg[ \frac{\sum_{t} \sigma_j(t) \exp (J\cdot \sigma(t) - \beta E_i(t))}{\sum_{t} \exp (J\cdot \sigma(t) - \beta E_i(t))} \bigg] \nonumber \\
& =\langle \delta \sigma_j \delta \sigma_k \rangle.
\end{align}
\item[(iii)]
\begin{align}
\frac{\partial}{\partial \beta} \bigg[ \frac{\partial^2 G}{\partial m_j \partial m_k} \bigg] &= \frac{\partial }{\partial \beta} [C^{-1}]_{jk}  \nonumber \\
&= - \sum_{\mu, \nu} [C^{-1}]_{j \mu} \frac{\partial C_{\mu \nu}}{\partial \beta} [C^{-1}]_{\nu k}  \nonumber \\
&= \sum_{\mu, \nu} [C^{-1}]_{j \mu} [C^{-1}]_{k \nu} \langle \delta E_i \delta \sigma_\mu \sigma_\nu \rangle. 
\end{align}
\item[(iv)]
\begin{align}
\frac{\partial^3 G}{\partial m_j \partial m_k \partial m_l} &= \frac{\partial }{\partial m_j} [C^{-1}]_{kl} 
= \sum_\lambda \frac{\partial J_\lambda}{\partial m_j} \frac{\partial}{J_\lambda} [C^{-1}]_{kl} \nonumber \\
&= - \sum_{\lambda, \mu, \nu} [C^{-1}]_{j \lambda} [C^{-1}]_{k \mu} \frac{\partial C_{\mu \nu}}{\partial J_\lambda} [C^{-1}]_{\nu l}  \nonumber \\
&= - \sum_{\lambda, \mu, \nu} [C^{-1}]_{j \lambda}[C^{-1}]_{k \mu} [C^{-1}]_{l \nu} \langle \delta \sigma_\lambda \delta \sigma_\mu \sigma_\nu \rangle.
\end{align}
\end{itemize}
Plugging these derivatives into Eq.~(\ref{eq:dGdb}), we obtain the following equation up to second order in $\delta m$:
\begin{align}
\langle \delta E_i \rangle' &= \sum_{j,k} \langle \delta E_i \delta \sigma_k \rangle^* [C^{-1}]^*_{kj} \langle \delta \sigma_j \rangle' \nonumber \\
&+ \frac{1}{2}  \sum_{j,k} \sum_{\mu, \nu} \langle \delta E_i \delta \sigma_{\mu} \sigma_{\nu} \rangle^*[C^{-1}]^*_{j\mu}[C^{-1}]^*_{k\nu} \langle \delta \sigma_j \rangle' \langle \delta \sigma_k \rangle' \nonumber \\
& - \frac{1}{2}  \sum_{j,k,l} \sum_{\lambda, \mu, \nu} \langle \delta E_i \delta \sigma_l \rangle^* \langle \delta \sigma_\lambda \delta \sigma_\mu \sigma_\nu \rangle^* \nonumber \\
& \phantom{MMMM} \times [C^{-1}]^*_{j\lambda} [C^{-1}]^*_{k\mu} [C^{-1}]^*_{l\nu} \langle \delta \sigma_j \rangle' \langle \delta \sigma_k \rangle', 
\end{align}
where we used the shorter notation: $\langle f \rangle' \equiv \langle f \rangle_{J, \beta=0}$, $\langle f \rangle^* \equiv \langle f \rangle_{J=0, \beta=0}$,
and $\langle \delta f \rangle' \equiv \langle f \rangle' - \langle f \rangle^*$.
Finally, we obtain the following relation:
\begin{equation}
\label{eq:H}
\langle \delta E_i \rangle' = \sum_j W_{ij}^* \langle \delta \sigma_j \rangle' + \frac{1}{2} \sum_{j,k} Q_{ijk}^* \langle \delta \sigma_j \rangle' \langle \delta \sigma_k \rangle',
\end{equation}
where 
\begin{equation}
W_{ij}^* \equiv \sum_{k} \langle \delta E_i \delta \sigma_k \rangle^* [C^{-1}]^*_{kj}
\end{equation}
and
\begin{align}
Q_{ijk}^* &\equiv \sum_{\mu, \nu} \langle \delta E_i \delta \sigma_{\mu} \sigma_{\nu} \rangle^*[C^{-1}]^*_{j\mu}[C^{-1}]^*_{k\nu} \nonumber \\
&- \sum_{l} \sum_{\lambda, \mu, \nu} \langle \delta E_i \delta \sigma_l \rangle^* \langle \delta \sigma_\lambda \delta \sigma_\mu \sigma_\nu \rangle^* [C^{-1}]^*_{j\lambda} [C^{-1}]^*_{k\mu} [C^{-1}]^*_{l\nu}. \nonumber \\
\end{align}
The second term in Eq.~(\ref{eq:H}) can be approximated as
\begin{align}
\langle \delta \sigma_j \rangle' \langle \delta \sigma_k \rangle' &= \big( \langle \sigma_j \rangle' - \langle \sigma_j \rangle^* \big) \big( \langle \sigma_k \rangle' - \langle \sigma_k \rangle^* \big) \nonumber \\
&\approx \langle \sigma_j \sigma_k \rangle' - \langle \sigma_j \sigma_k \rangle^* \nonumber \\
& - \langle \sigma_j \rangle^* \big( \langle \sigma_k \rangle' - \langle \sigma_k \rangle^* \big) - \langle \sigma_k \rangle^* \big( \langle \sigma_j \rangle' - \langle \sigma_j \rangle^* \big) \nonumber \\
&= \langle \delta (\sigma_j \sigma_k) \rangle' - \langle \sigma_j \rangle^* \langle \delta \sigma_k \rangle' - \langle \sigma_k \rangle^* \langle \delta \sigma_j \rangle',
\end{align}
where the second line assumes a negligible correlation between $\sigma_j$ and $\sigma_k$:
$\langle \sigma_j \sigma_k \rangle \approx \langle \sigma_j \rangle \langle \sigma_k \rangle$.
Then, with the Rao-Blackwell conditional expectation update $H_i(m)^{\textrm{new}} \leftarrow \langle E_i \rangle_{J(m^*)}$, Eq.~(\ref{eq:H}) implies 
\begin{equation}
H_i = \sum_j \bigg(W_{ij}^* - \sum_k Q_{ijk}^* \langle \sigma_k \rangle^* \bigg) \sigma_j + \frac{1}{2} \sum_{j,k} Q_{ijk}^* \sigma_j \sigma_k,
\end{equation}
where we used $Q_{ijk}=Q_{ikj}$.
This formalism allows one to infer the linear and quadratic relations between $H_i$ and $\sigma$.

\bigskip
\subsection*{SI Text 2: Review on the mean-field methods for the kinetic Ising model}

\subsubsection*{Maximum likelihood estimation (MLE)}
The kinetic Ising model updates spins with the conditional probability,
\begin{equation}
\label{eq:condprob}
P(\sigma_i(t+1)=\pm 1|\sigma(t)) = \frac{\exp(\pm H_i(\sigma(t)))}{\exp(H_i(\sigma(t))) + \exp(-H_i(\sigma(t)))},
\end{equation}
where $H_i(\sigma(t)) = \sum_j W_{ij} \sigma_j(t)$. Then, the expectation value of $\sigma_i(t+1)$ given $\sigma(t)$ becomes
\begin{align}
\langle \langle \sigma_i(t+1)) \rangle \rangle_{\sigma(t)} &= \sum_{\rho = \{1, -1\}} \rho P(\sigma_i(t+1)=\rho|\sigma(t)) \nonumber \\
&= \tanh (H_i(\sigma(t))).
\end{align}
Given $N$-dimensional time-series data $\sigma(t)$ with length $L$, the data likelihood is defined as
\begin{equation}
{\cal{P}} = \prod_{t=1}^{L-1}\prod_{i=1}^{N} P(\sigma_i(t+1)|\sigma(t)).
\end{equation}
Using MLE, one can optimize $W_{ij}$ to increase $\log \cal{P}$:
\begin{equation}
W_{ij}^{\textrm{new}} = W_{ij} + \frac{\alpha}{L-1} \frac{\partial \log {\cal{P}} }{\partial W_{ij}}
\end{equation}
with a learning rate $\alpha$~\cite{Zeng2013}. 
Here, one can calculate the gradient with Eq.~(\ref{eq:condprob}),
\begin{equation}
\label{eq:gradient}
\frac{\partial \log {\cal{P}} }{\partial W_{ij}} = \sum_{t=1}^{L-1} \bigg( \sigma_i(t+1) \sigma_j(t) - \tanh \big(H_i(\sigma(t)) \big) \sigma_j(t) \bigg).
\end{equation}

\subsubsection*{Na{\"i}ve mean-field approximation (nMF)}
The maximum condition of the log-likelihood ($\partial \log {\cal{P}}/\partial W_{ij}$=0) in Eq.~(\ref{eq:gradient}) gives 
\begin{equation}
\label{eq:maxlog}
\sum_{t=1}^{L-1} \sigma_i(t+1) \sigma_j(t) = \sum_{t=1}^{L-1} \tanh \big(H_i(\sigma(t)) \big) \sigma_j(t)
\end{equation}
with $H_i(\sigma(t))=\sum_k W_{ik} \sigma_k(t)$.
For a mean-field approximation, spin activities are represented by the mean field activity plus its residual: $\sigma_i(t) = m_i + \delta \sigma_i(t)$.
Then, using the Taylor expansion, one can approximate $\tanh \big(H_i(\sigma(t)) \big) \approx \tanh (g_i) + \big( 1 - \tanh^2 (g_i ) \big) \sum_k W_{ik} \delta \sigma_k(t)$ with $g_i = \sum_k W_{ik} m_k$.
The zeroth-order expectation of $\langle \langle \sigma_i(t+1) \rangle \rangle_{\sigma(t)} \approx \tanh (g_i)$ gives the self-consistent equation
\begin{equation}
\label{eq:nMF}
m_i = \tanh \big( \sum_k W_{ik} m_k \big).
\end{equation}
Then, using the mean-field approximation, Eq.~(\ref{eq:maxlog}) becomes
\begin{align}
&\sum_t \big( m_i + \delta \sigma_i(t+1) \big) \big( m_j + \delta \sigma_j(t) \big) \nonumber \\
&= \sum_t \big( m_i + (1-m_i^2) \sum_k W_{ik} \delta \sigma_k(t) \big) \big( m_j + \delta \sigma_j(t) \big)
\end{align}
Given the data with length $L$,
\begin{align}
&\frac{1}{L-1} \sum_{t=1}^{L-1} \delta \sigma_i(t+1) \delta \sigma_j(t) \nonumber \\
&= (1-m_i^2) \sum_k W_{ik} \frac{1}{L-1} \sum_{t=1}^{L-1} \delta \sigma_k(t) \delta \sigma_j(t). 
\end{align}
One can also derive this equation from $\delta \sigma_i (t+1) = (\partial m_i/\partial m_k) \delta \sigma_k(t)$ with Eq.~(\ref{eq:nMF}).
The equality gives a matrix equation to infer
\begin{equation}
W_{\text{nMF}}=A_{\text{nMF}}^{-1} B C^{-1},
\end{equation}
where $[A_{\text{nMF}}]_{ij} = (1-m_i^2) \delta_{ij}$ is a diagonal matrix; $B_{ij} = \langle \delta \sigma_i(t+1) \delta \sigma_j(t) \rangle$ is a time-delayed correlation; and the covariance matrix $C_{ij} = \langle \delta \sigma_i(t) \delta \sigma_j(t) \rangle$ is an equal-time correlation~\cite{Roudi2011}.

\subsubsection*{Thousless-Anderson-Palmer mean-field approximation (TAP)} 
Compared to nMF, TAP considers the second-order correction of the Onsager's reaction term:
\begin{align}
\label{eq:TAP}
&\langle \sigma_i(t+1) \rangle = \big\langle \tanh \big( \sum_k W_{ik} \sigma_k(t) \big) \big\rangle \nonumber \\
&\approx \tanh(g_i) + \frac{1}{2} \bigg[ \frac{\partial^2 \tanh (x)}{\partial x^2} \bigg]_{x=g_i} \langle \delta g_i^2 \rangle \nonumber \\
&\approx \tanh(g_i) - \big( 1 - \tanh^2(g_i) \big) \tanh (g_i) \sum_l W_{il}^2 (1-m_l^2) 
\end{align}
with $g_i \equiv \sum_k W_{ik} m_k$, $\delta g_i \equiv \sum_k W_{ik} \delta \sigma_k(t)$, and
\begin{align}
\label{eq:variance}
\langle \delta g_i^2 \rangle &= \sum_{k,l} W_{ik} W_{il} \langle \delta \sigma_k \delta \sigma_l \rangle = \sum_l W_{il}^2 \langle \delta \sigma_l^2 \rangle\nonumber \\
&= \sum_l W_{il}^2 \langle (\sigma_l - m_l)^2 \rangle = \sum_l W_{il}^2  (1-m_l^2)
\end{align}
under the assumption of the negligible correlation between $\sigma_k$ and $\sigma_l$: $\langle \delta \sigma_k \delta \sigma_l \rangle\approx0$ for $k \neq l$.
The correction gives a refined self-consistent equation
\begin{equation}
\label{eq:TAP2}
m_i = \tanh \big( \sum_k W_{ik} m_k - m_i \sum_l W_{il}^2 (1-m_l^2) \big).
\end{equation}
Then, using $\delta \sigma_i (t+1) = (\partial m_i/\partial m_k) \delta \sigma_k(t)$, one can derive
\begin{equation}
\delta \sigma_i(t+1) = (1-m_i^2) (1-F_i) \sum_k W_{ik} \delta \sigma_k(t)
\end{equation}
with $F_i \equiv (1-m_i^2) \sum_l W_{il}^2 (1-m_l^2)$. This leads to
\begin{equation}
\langle \delta \sigma_i(t+1) \delta \sigma_j(t) \rangle = (1-m_i^2)(1-F_i) \sum_k W_{ik} \langle \delta \sigma_k(t) \delta \sigma_j(t) \rangle.
\end{equation}
Therefore, one obtains the TAP estimates
\begin{equation}
W_{\text{TAP}} = (1-F_i)^{-1} W_{\text{nMF}}.
\end{equation}
Here, one can obtain $F_i$ as a solution of the self-consistent equation~\cite{Roudi2011}:
\begin{equation}
F_i (1-F_i)^2 = (1-m_i^2) \sum_l [W_{\text{nMF}}]_{il}^2 (1-m_l^2). 
\end{equation}

\subsubsection*{Exact mean-field approximation (eMF)} 
For random $W_{ik}$ with a large number $N$ of spin components, it is a reasonable assumption that $H_i = \sum_{k=1}^N W_{ik} \sigma_k$ follows a Gaussian distribution with a mean $g_i = \sum_k W_{ik} m_k$ and a variance $\Delta_i = \langle \delta g_i^2 \rangle = \sum_l W_{il}(1-m_l^2)$ in Eq.~(\ref{eq:variance}):
\begin{equation}
\langle \delta \sigma_i(t+1) \rangle = \int_{-\infty}^{\infty} dx \frac{e^{-x^2}}{\sqrt{2 \pi}} \tanh (g_i + x \sqrt{\Delta_i}).
\end{equation}
Here, the zeroth-order and second-order Taylor expansion of $\tanh (g_i + x \sqrt{\Delta_i})$ with respect to $x$ give the nMF and TAP solutions in Eqs.~(\ref{eq:nMF}) and (\ref{eq:TAP2}).
The multi-variable $x \equiv \delta g_i$ and $y \equiv \delta g_j$ may also follow a Gaussian distribution:
\begin{equation}
P(x, y) = \frac{1}{2 \pi \sqrt{\Delta_i \Delta_j}} \exp \bigg[ - \frac{x^2}{2 \Delta_i} - \frac{y^2}{2 \Delta_j} + \Delta_{ij} \frac{xy}{\Delta_i \Delta_j}\bigg],
\end{equation}
where the covariance $\Delta_{ij}$ is defined as
\begin{eqnarray}
\Delta_{ij} &\equiv& \langle \delta g_i \delta g_j \rangle = \bigg\langle \sum_k W_{ik} \delta \sigma_k \sum_l W_{jl} \delta \sigma_l \bigg\rangle\nonumber \\
&=&\sum_{k,l} W_{ik} \langle \delta \sigma_k \delta \sigma_l \rangle W_{lj}^\top = [W C W^\top]_{ij}.
\end{eqnarray}
Then, the time-delayed correlation matrix $B$ can be approximated as
\begin{align}
B_{ik} &= \langle \delta \sigma_i(t+1) \delta \sigma_k(t) \rangle \nonumber \\
&= \langle \sigma_i(t+1) \sigma_k(t) \rangle - \langle \sigma_i(t+1) \rangle \langle \sigma_k(t) \rangle \nonumber \\
&= \langle \sigma_k(t) \tanh (g_i + \delta g_i(t)) \rangle - \langle \sigma_k(t) \rangle \langle \tanh (g_i + \delta g_i(t)) \rangle.
\end{align}
Using $B$, one can derive $BW^\top=AWCW^\top$ as follows:
\begin{align}
\sum_k W_{jk} B_{ik} &= \big\langle \sum_k W_{jk} \sigma_k(t) \tanh (g_i + \delta g_i(t) \big\rangle \nonumber \\
& \phantom{MM} - \big\langle \sum_k W_{jk} \sigma_k(t) \big\rangle \big\langle \tanh(g_i + \delta g_i(t)) \big\rangle \nonumber \\
&= \langle (g_j + \delta g_j(t)) \tanh (g_i + \delta g_i(t)) \rangle \nonumber \\
&  \phantom{MM} - \langle g_j + \delta g_j(t) \rangle \langle \tanh (g_i + \delta g_i(t)) \rangle \nonumber \\
&= \langle \delta g_j \tanh (g_i + \delta g_i) \rangle \nonumber \\
&= \int_{-\infty}^{\infty} \frac{dx dy}{2 \pi \sqrt{\Delta_i \Delta_j}} y \tanh(g_i + x) \nonumber \\
& \phantom{MMMMM} \times \exp \bigg[-\frac{x^2}{2 \Delta_i} - \frac{y^2}{2 \Delta_j} + \Delta_{ij} \frac{xy}{\Delta_i \Delta_j} \bigg]  \nonumber \\
&\approx \frac{\Delta_{ij}}{\Delta_i \Delta_j} \int_{-\infty}^{\infty} \frac{dx}{\sqrt{2 \pi \Delta_i}} \frac{dy}{\sqrt{2 \pi \Delta_j}} xy^2 \tanh(g_i + x) \nonumber \\
& \phantom{MMMMM} \times \exp \bigg[-\frac{x^2}{2 \Delta_i} - \frac{y^2}{2 \Delta_j} \bigg] \nonumber \\
&= \frac{\Delta_{ij}}{\Delta_i} \int_{-\infty}^{\infty} \frac{dx}{\sqrt{2 \pi \Delta_i}} x \tanh(g_i + x) \exp \bigg[-\frac{x^2}{2 \Delta_i} \bigg] \nonumber \\
&= \Delta_{ij} \int_{-\infty}^{\infty} \frac{dx}{\sqrt{2 \pi \Delta_i}} \exp \bigg[-\frac{x^2}{2 \Delta_i} \bigg] \big( 1 - \tanh^2 (g_i + x) \big) \nonumber \\
& = [W C W^\top]_{ij} a_i.
\end{align}
This equation gives
\begin{equation}
W_{\text{eMF}} = A_{\text{eMF}}^{-1} B C^{-1},
\end{equation}
where $[A_{\text{eMF}}]_{ij} = a_i \delta_{ij}$ is a diagonal matrix.
In practice, one can obtain $W_{\text{eMF}}$ with the following iterations~\cite{Mezard2011}:
\begin{itemize}
\item[(i)] Calculate $\Delta_i$ (Gguess $\Delta_i$ for the first round):
\begin{equation}
\Delta_i = \frac{1}{a_i^2} \sum_j [BC^{-1}]_{ij}^2 (1-m_j^2).
\end{equation}
\item[(ii)] Find $g_i$ as a solution for the following integral equation:
\begin{equation}
m_i = \int_{-\infty}^{\infty} \frac{dx}{\sqrt{2 \pi}} \exp \bigg[-\frac{x^2}{2} \bigg] \tanh (g_i + x \sqrt{\Delta_i}).
\end{equation} 
\item[(iii)] Calculate $a_i$ given $g_i$ and $\Delta_i$:
\begin{equation}
a_i = \int_{-\infty}^{\infty} \frac{dx}{\sqrt{2 \pi}} \exp \bigg[-\frac{x^2}{2} \bigg] \big( 1 - \tanh^2 (g_i + x \sqrt{\Delta_i}) \big).
\end{equation}
\end{itemize}

\bigskip
\subsection*{SI Text 3: Neuronal data processing}
In the original data, neuron $i$ is defined as ``active'' ($\sigma_i(t) = 1$), if the neuron fires at least once during the time window $[t, t+\delta t]$, otherwise ``silent'' ($\sigma_i(t)=-1)$ (Fig.~\ref{fig:figS2}, upper).
To suppress the dependency of the time interval $\delta t$ for the activity definition, we used a moving average of activities.
We examined the past five and future five activities of neuron $i$, and redefined $\sigma_i(t)=1$, if neuron $i$ emitted at least one spike in the time window, otherwise $\sigma_i(t)=-1$ (Fig.~\ref{fig:figS2}, lower).
Since neurons may have a refractory period that prevents consecutive spikes after emitting a spike~\cite{Kara2000}, the moving average can also help infer the genuine interaction between neurons by reducing the effect of the refractory period.

For the estimation of $W_{ij}$ and $H_i^{\textrm{ext}}$, we estimated $W_{ij}$ first with $H_i^{\textrm{ext}}=0$, and then estimated $W_{ij}$ and $H_i^{\textrm{ext}}$ together because $H_{i}^{\textrm{ext}}$ turned out to be quite large compared to $W_{ij}$.
These training procedures were repeated for $20$ times.

\begin{figure*}[h]
\centering
\includegraphics[width=9cm]{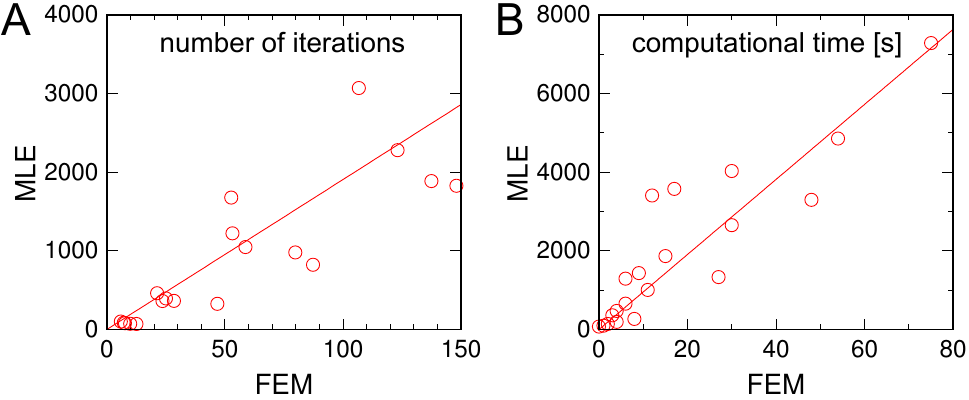}
\caption{ \label{fig:figS1} Efficiency of inference. Number of iterations per spin (A) and real computational time (B) by using MLE versus FEM for various coupling strengths, $g$ from $1$ to $4$ and number of observed configurations, $L/N^2$ from $0.2$ to $1$. A system size $N=100$ is used. A learning rate $\alpha=1$ is used for MLE.
}
\end{figure*}


\begin{figure*}
\centering
\includegraphics[width=9cm]{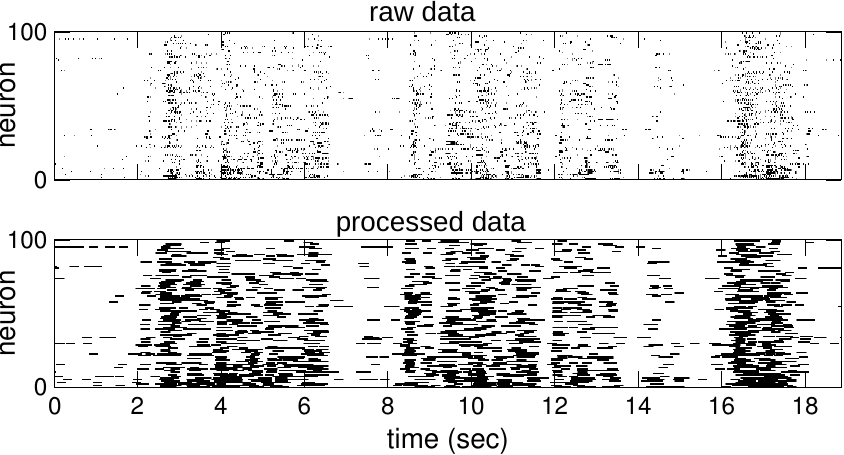}
\caption{\label{fig:figS2} Data processing. Rasters of 100 neuronal activities from raw (upper) and processed (lower) data are plotted with black dots representing spikes, white dots representing queiscence.  
}
\end{figure*}

\begin{figure*}[h]
\centering
\includegraphics[width=15cm]{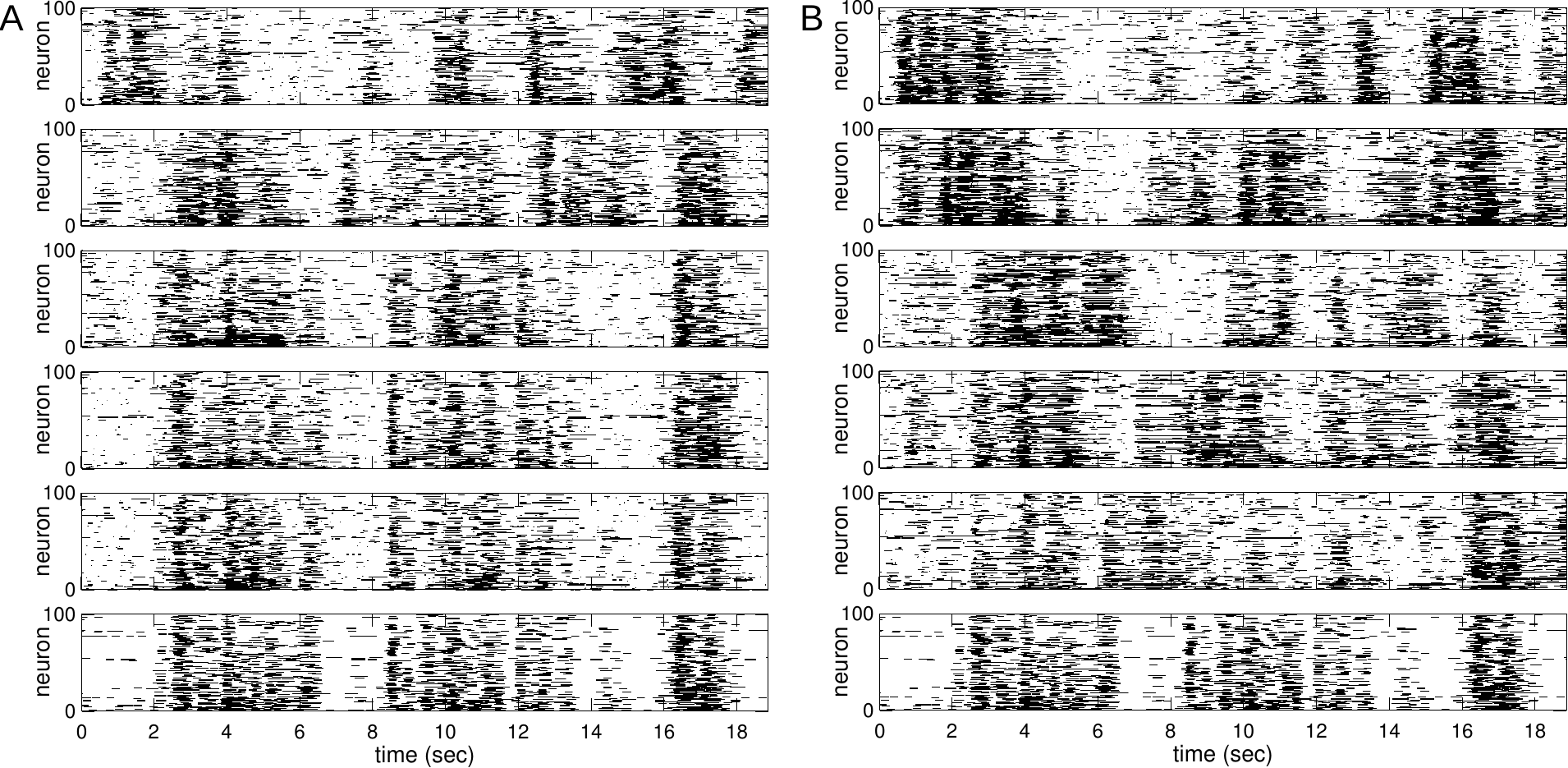}
\caption{\label{fig:figS3} Reconstruction of neuronal activity. The raster of neuronal activities is recovered using our method, with large-$|W_{ij}|$-based selection (A) and under random selection (B) with various numbers of input neurons, from top to bottom: $1$, $2$, $8$, $10$, $12$. The actual raster is also shown on the bottom of each column for comparison.
}
\end{figure*}

\begin{figure*}[h]
\centering
\includegraphics[width=9cm]{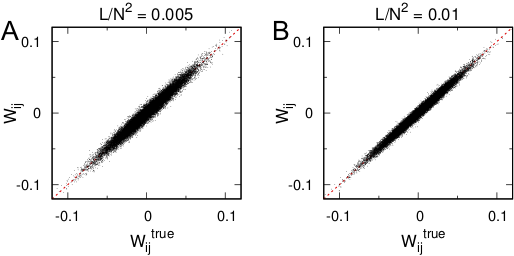}
\caption{\label{fig:figS4} Inference of coupling strength in large system size, $N=5000$. Predicted couplings versus actual couplings for $L/N^2 = 0.005$ (A) and $0.01$ (B). The actual coupling strengths are normally-distributed with $g=2$. The computation time for this simulation is approximately 4 days and 8 days, respectively, for $L/N^2 = 0.005$ and $0.01$ on a 2.30 GHz processor.
}
\end{figure*}




\end{document}